\documentclass[aps,prc,preprint,tightenlines,groupedaddress,
nofootinbib,showpacs,preprintnumbers,amsmath,amssymb,floatfix,
superscriptaddress]{revtex4} 
\usepackage{graphicx} 
\usepackage{dcolumn} 
\usepackage{bm}
\usepackage{amsmath}
\def\subF #1{#1_{\lower2pt\hbox{$\scriptstyle{\rm F}$}}}
\usepackage{float}

\begin{document} 
\title{Compact Objects for Everyone I: White Dwarf Stars}
\author{C.B.~Jackson, J.~Taruna, S.L.~Pouliot, B.W.~Ellison,
        D.D.~Lee, and J.~Piekarewicz}
\affiliation{Department of Physics, Florida State University, 
             Tallahassee, FL 32306, USA}
\date{\today}
\begin{abstract}
Based upon previous discussions on the structure of compact stars
geared towards undergraduate physics students, a real experiment
involving two upper-level undergraduate physics students, a beginning
physics graduate, and two advanced graduate students was conducted.  A
recent addition to the physics curriculum at Florida State University,
{\it The Physics of Stars}, sparked quite a few students' interests in
the subject matter involving stellar structure. This, coupled with
{\it Stars and Statistical Physics} by Balian and
Blaizot~\cite{Bla99_AMJ67} and {\it Neutron Stars for Undergraduates}
by Silbar and Reddy~\cite{Sil04_AJP72}, is the cornerstone of this
small research group who tackled solving the structure equations for
compact objects in the Summer of 2004. Through the use of a simple
finite-difference algorithm coupled to Microsoft Excel and Maple, 
solutions to the equations for stellar structure are presented in the 
Newtonian regime appropriate to the physics of white dwarf stars.
\end{abstract}
%
%
\maketitle

\section{Overview of the project}
It is the central tenet of the present ``experiment'' that advances in
both algorithms and computer architecture bring the once-challenging
problem of the structure of compact objects within the reach of
beginning undergraduate students --- and even high-school students.
After a brief historical review in Sec.~\ref{Historical}, a synopsis
of stellar evolution is presented in Sec.~\ref{StellarEvolution} that
culminates with a detailed description of the physics of collapsed
stars. Following this background information, the equations of
hydrostatic equilibrium for Newtonian stars are derived in
Sec.~\ref{NewtonianStars}.  In particular, the need for an equation of
state and the enormous advantage of {\it scaling} the equations are
emphasized. The section concludes with the presentation of {\it
mass-vs-radius} relationships for white dwarf stars obtained using
both Excel and Maple. The special role played by special relativity
for the existence of a limiting mass, the {\it Chandrasekhar limit}, 
is strongly emphasized.  Finally, a summary and some concluding remarks
are offered in Sec.~\ref{Conclusions}.

\section{Historical perspective}
\label{Historical}

The story of collapsed stars starts in earnest with Subrahmanyan
Chandrasekhar in the early 1930's. As a young man of 20, he was
embarking from his native India to Cambridge University to start life
as a graduate student under Fowler's supervision. By 1926 Fowler had
already explained the structure of white dwarf stars by using the
electron degenerate pressure only a few months after the formulation
of the Fermi-Dirac statistics~\cite{Fow26_RAS87}.  However,
Chandrasekhar noticed a critical ingredient missing from Fowler's
analysis: {\it special relativity}~\cite{Cha84_RMP56}.  Chandrasekhar
discovered that as the density of the star increases and the momentum
of the electrons 
$p$ becomes comparable with and later exceeds $mc$, where $m$ is the
rest mass and $c$ the speed of light, then their ability to support
the star against gravity weakens.  Chandrasekhar concluded that stars
with masses above $M_{Ch} \equiv 1.44 M_{\odot}$ ($M_{\odot}$ a solar
mass) cannot cool down but will continue to contract and heat up. This
limiting mass $M_{Ch}$ is fittingly known as the Chandrasekhar
mass. However, it was not all smooth sailing for Chandrasekhar. One
pre-eminent figure in the field, Sir Arthur Eddington, opposed
Chandrasekhar publicly and privately with great vigor, even claiming
 ---  by arguments found generally difficult to follow --- that the
non-relativistic pressure-density relation should be used at all
densities.
 
Why wouldn't Chandrasekhar silence his critics by revealing the
ultimate fate of a heavy ($M\!>\!M_{\rm Ch}$) star?  After all, we now
know that such a star will collapse into a neutron star, or if too
heavy into a black hole. Unfortunately for Chandrasekhar, at the time
of his ground-breaking discovery it was impossible for him (or for
anyone else for that matter) to have predicted the existence of
neutron stars. It was one year later in 1932 that Chadwick proved the
existence of the neutron~\cite{Chad32_PRSL136}. From that point on
things developed very quickly, culminating with the 1933 proposal by
Baade and Zwicky that supernovae are created by the collapse of a
``normal'' star to form a neutron
star~\cite{Baad34_PR46}. Chandrasekhar was thus vindicated and awarded
the Nobel Prize in Physics in 1983 for his lifetime contributions to
the {\it physical processes of importance to the structure and
evolution of the stars}.

The process of discovery of these two classes of compact objects
(white dwarf and neutron stars) is radically different. In the case of
white dwarf stars, observation predated Fowler's theoretical
explanation by more than 10 years. Indeed, in 1915 at the Mount Wilson
Observatory in California a group of astronomers headed by Walter
Sydney Adams discovered that Sirius B --- the companion of the brightest
star in the night sky --- was a white-dwarf star, the first to be
discovered.  On the other hand, it took more than 30 years since the
bold prediction by Baade and Zwicky~\cite{Baad34_PR46} to discover
neutron stars. As is often the case in science, this discovery marks 
one of the greatest examples of serendipity. Antony Hewish and his
team at the Cavendish Laboratory built a radio telescope to study some
of the most energetic quasi-stellar objects in the universe ({\it
quasars}). Among the members of the team was a young research student 
by the name of Jocelyn Bell~\cite{Bel04_Sci304}.  Shortly after the
telescope started gathering data in 1967, Bell observed a signal of
seemingly unknown origin (a ``scruff''). Particularly puzzling was
that the signal showed at a remarkably precise pulse rate of 1.33
seconds~\cite{Hew68_Nat217}.  So unexpected was the signal, that after
a month of futile attempts at understanding it, it was dubbed the
``Little Green Men.'' However, by the beginning of 1968 Hewish, Bell,
and collaborators had found three additional pulsating sources of
radio waves, or {\it pulsars}~\cite{Pil68_Nat218}.  The final
explanation of these enigmatic sources was due to Gold shortly after
Hewish and Bell published their
findings~\cite{Gold68_Nat218,Gold69_Nat221}.  Gold suggested that the
radio signals were due to rapidly rotating neutron stars that rather
than emitting pulses of radiation, could emit a steady radio signal
that it swept around in circles. When the pulsar ``lighthouse'' was
pointing in the direction of the telescope, the signal will indeed
show up as the short ``pulses'' that Bell had discovered. It is not
our intention to present here a comprehensive review of either the
history or the fascinating phenomena behind pulsars. For a recent
account see Ref.~\cite{Science304}.

\section{Stellar Evolution -- a Synopsis}
\label{StellarEvolution}
Stellar objects are dynamical systems involving a symbiotic
relationship between matter and radiation creating enough pressure to
oppose gravitational contraction. Thermonuclear fusion,
the thermally-induced combining of nuclei as they tunnel through the 
Coulomb barrier, is initially responsible for supporting stars against 
gravitational contraction. The ultimate fate of the star depends 
upon its remaining mass once thermonuclear fusion can no longer provide 
the pressure required to counteract gravity. 

Thermonuclear fusion drives stars through many stages of combustion; the 
hot center of the star allows hydrogen to fuse into helium. Once the core 
has burned all available hydrogen, it will contract until another source 
of support becomes available. As the core contracts and heats, transforming 
gravitational energy into kinetic (or thermal) energy, the burning of the 
helium ashes begins. For stars to burn heavier elements, higher temperatures 
are necessary to overcome the increasing Coulomb repulsion and allow fusion 
through quantum-mechanical tunneling. Thermonuclear burning continues until 
the formation of an iron core. Once iron --- the most stable of nuclei --- is 
reached, fusion becomes an endothermic process. However, combustion to iron 
is only possible for the most massive of stars. When thermonuclear 
fusion can no longer support the star against gravitational collapse, 
either because they are not massive enough (like our Sun) or 
because they have developed an iron core, the star dies and a compact 
object is ultimately formed. The three final possible stages a star 
can take is a white dwarf, a neutron star, or a black hole. In this
our first contribution --- one that should be accessible to motivated 
high-school students --- we focus exclusively on the physics of white 
dwarf stars. The fascinating topic of neutron stars requires the use 
of {\it general relativity} and is therefore reserved for a more 
advanced forthcoming publication.

Our Sun will die as a white dwarf star once all of the hydrogen and
helium in the core has been burned. Towards the final stages of
burning, the star will expand and expel most of the outer matter to
create a planetary nebula.  
At the beginning, the non-degenerate core contracts and heats up
through conversion of gravitational energy into thermal kinetic energy.
However, at some point the Fermi pressure of the degenerate electrons
begins to dominate, the contraction is slowed up, and the core becomes
a compact object known as a white dwarf, cooling steadily towards the
ultimate cold, dark, static black dwarf state. On the other hand,
neutron stars
result from one of the most cataclysmic events in the universe, the
death of a star with a mass much greater than that of our
Sun. Electrons in these stars behave ultra-relativistically, and as
pointed out by Chandrasekhar,
hydrostatic equilibrium as a cold body becomes impossible to
achieve when $M\!>\!M_{Ch}$. However, during the collapse of the core,
a supernovae shock develops ejecting most of the mass of the
star into the interstellar space and leaving behind an extremely dense
core --- the neutron star.  As the star collapses, it becomes
energetically favorable for electrons to be captured by protons,
making neutrons and neutrinos. The neutrinos carry away 99\% of the
gravitational binding energy of the compact object, leaving neutrons
behind to support the star against further collapse.  The pressure
provided by the degenerate neutrons, like degenerate electron pressure
for white dwarf stars, has a limit on the mass it can bear. Beyond
this limiting mass, no source of pressure exists that can prevent
gravitational contraction. If such is the case, then the star will
continue to collapse into an object of zero radius: a black hole.
There is a large number of excellent textbooks on the birth, life, and
death of stars. The following are some references used in this
work~\cite{Wei72_JW,Mis73_WH,Sh83_JW,Thor94_WWN,Gle00_SV,Phi02_JW}.

\subsection{The Physics of Collapsed Stars}
It is a remarkable fact that quantum mechanics and special
relativity --- both theories perceived as of the very small --- play such
a crucial role in the dynamics of stars, the former in preventing
low-mass stars from collapsing into black holes, while the latter in
driving the collapse of massive stars. In this research experiment we
limit ourselves to the study of white dwarf (or Newtonian) stars. 
Further, we assume that white dwarf stars are ``cold'', spherically 
symmetric, non-rotating objects in hydrostatic equilibrium.

In dealing with white dwarf stars, the system may be approximated as a
plasma containing positively charged nuclei and electrons, with the
nuclei providing (almost) all the mass and none of the pressure and the
electrons providing all the pressure and none of the mass. This state
of matter corresponds to a gas that is electrically neutral on a
global scale, but locally composed of positively charged ions (nuclei)
and the negatively charged electrons. 
Note that even in a
zero-temperature, black dwarf state, the matter is effectively
ionized. For a free atom, the Heisenberg Uncertainty Relation ensures
that in the state of minimum total energy  --- kinetic plus electric
potential energy --- the electrons occupy a finite volume of
dimension given by the Fermi-Thomas distance, the analogue for
more massive atoms of the Bohr radius for hydrogen. The density of
a collapsed star is so high that the mean distance between nuclei
is less than the Fermi-Thomas distance: the matter is
`pressure-ionized', with the electrons forming an effectively free
gas. However, all identical fermions (such as electrons, protons
and neutrons) obey Fermi-Dirac statistics in that the occupation
of states is governed by the Pauli Exclusion Principle --- no two
fermions can exist in the same quantum state.
For a zero temperature Fermi gas,
all available electron states below the {\it Fermi energy} are filled,
while the rest are empty. The Fermi energy is determined solely by the
electron number density and rest mass. For a compact object such as a
white dwarf star, the number density is very high and so is the Fermi
energy. Typical electron Fermi energies are of the order of 1 MeV
which correspond to a {\it Fermi temperature} of 
$T_{\rm F}\!\simeq\!10^{10}$~K.  As the temperature of the system is
increased (say from $T\!=\!0$ to $T\!=\!10^{6}$~K
or more, as estimated for white dwarf interiors),
electrons try to jump to a state higher in energy by an amount of the
order of $k_{\rm B}T$ but fail, as most of these transitions are Pauli
blocked.  Only those high-energy electrons that are within $k_{\rm
B}T\!\simeq\!100$~eV from the Fermi surface can make the transition,
but those represent a tiny fraction ($T/T_{\rm F}\!\simeq\!10^{-4}$)
of the electrons in the star. Hence, for the purpose of computing the
pressure of the system, it is extremely accurate  ---  to 1 part in
$10^{4}$  ---  to describe the electrons as a Fermi gas at zero
temperature.

\section{Newtonian Stars}
\label{NewtonianStars}
Let us start by addressing Newtonian stars. For these stars we 
assume that corrections due to Einstein's greatest triumph --- the 
{\it Theory of General Relativity} --- may be safely ignored. White 
dwarf stars, with escape velocities of only 3\% of the speed of 
light, fall into this category. Not so neutron stars --- typical 
escape velocities of half the speed of light cause extreme 
sensitivity to these corrections.

\begin{figure}[h]
 \includegraphics[height=8cm,angle=0]{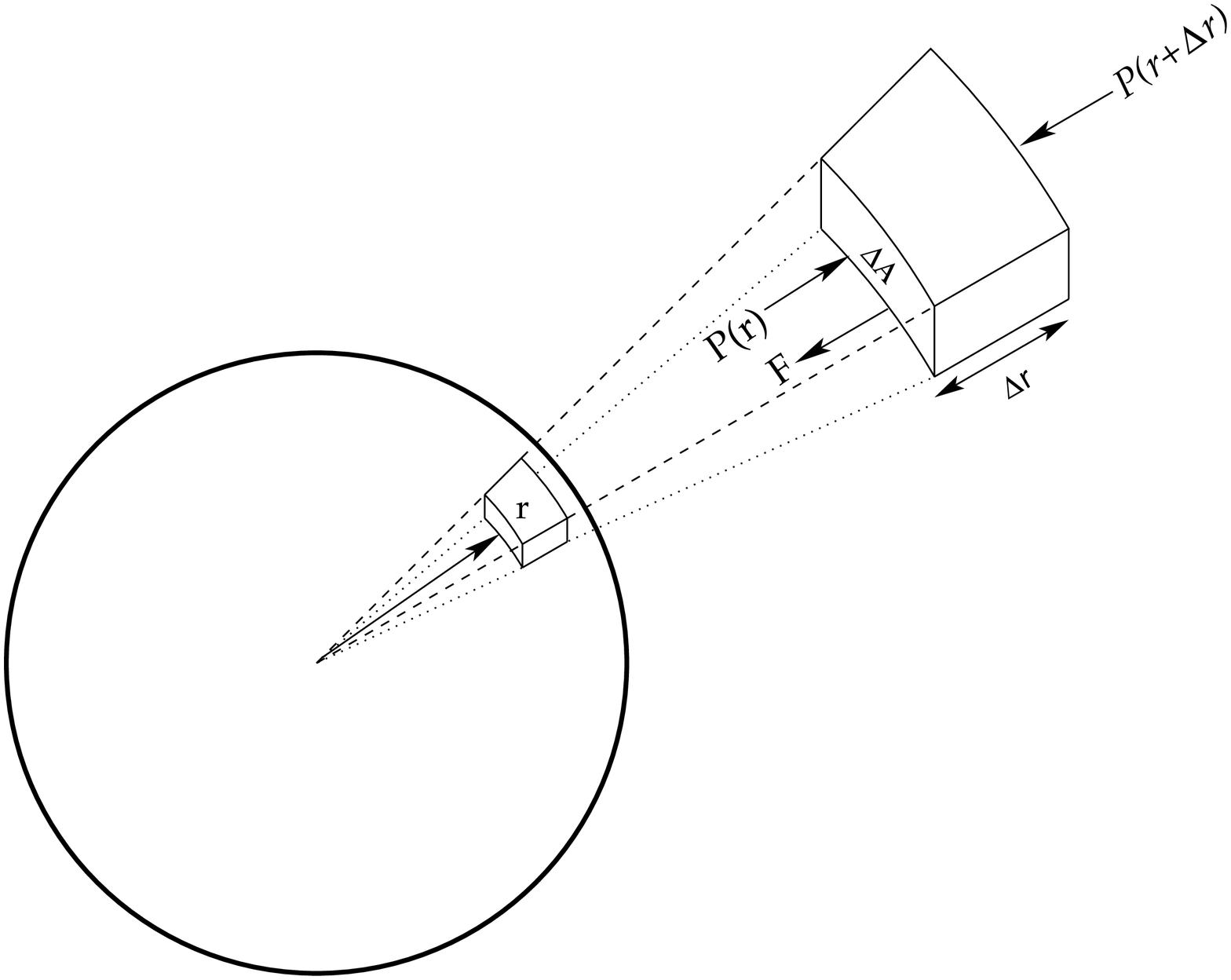}
 \caption{The radial force acting on a small mass element a 
    distance r from the center of the star.}
 \label{fig1}
\end{figure}

We start by considering the radial force acting on a small mass element 
($\Delta m\!=\!\rho(r)\Delta V$) located at a distance $r$ from the 
center of the star (see Fig.~\ref{fig1}):
\begin{equation}
  F_{r}=-\frac{GM(r)\Delta m}{r^2}-P(r+\Delta r)\Delta A
        +P(r)\Delta A=\Delta m \frac{d^{2}r}{dt^{2}} \;.
 \label{Force}
\end{equation}
Here $\rho(r)$ is the mass density of the star, $M(r)$ denotes the 
{\it enclosed mass} within a radius $r$, and $P$ is the pressure. 
Expanding the above equation to lowest order in $\Delta r$ one obtains
\begin{equation}
   -\frac{GM(r)\rho(r)}{r^2}-\frac{dP}{dr}=
    \rho(r)\frac{d^{2}r}{dt^{2}} \;.
 \label{PreHydroEq}
\end{equation}
Assuming {\it hydrostatic equilibrium} 
($\ddot{r}\!=\!\dot{r}\!\equiv\!0$), one arrives at the fundamental equations 
describing the structure of Newtonian stars.  That is,
\begin{subequations}
\begin{eqnarray}
   &&\frac{dP}{dr}=-\frac{GM(r)\rho(r)}{r^2}\;, 
   \quad P(r\!=\!0)\equiv P_{\rm c} \;; 
 \label{HydroEqsa} \\
   &&\frac{dM}{dr}=+4\pi r^{2}\rho(r) \;, 
   \hspace{0.32in} M(r\!=\!0)\equiv 0 \;, 
 \label{HydroEqsb}
\end{eqnarray}
 \label{HydroEqs}
\end{subequations}
where Eq.~(\ref{HydroEqsb}) defines the enclosed mass.

It is simple to see that in hydrostatic equilibrium, the pressure of
the star is a decreasing (or at least not increasing) function of $r$;
otherwise the star collapses. Note that the radius of the star $R$ is 
defined as the value of $r$ at which the pressure goes to zero,
{\it i.e.,} $P(R)\!=\!0$. Similarly, the mass of the star corresponds
to the value of the enclosed mass at $r\!=\!R$, when $M\!=\!M(R)$.

\subsection{Equation of State}
\label{EquationofState}
The above set of equations, together with their associated boundary
conditions, must be completed by an equation of state (EoS), namely a
relation $P = P(\rho)$ between the density and pressure. For simplicity,
we limit ourselves
to the EoS of a zero-temperature Fermi gas composed of constituents
({\it e.g.,} electrons or neutrons) having a rest mass $m$. The main
assumption behind the Fermi gas hypothesis is that no correlations (or
interactions) are relevant to the system other than those generated by
the Pauli exclusion principle. For some standard references on the
equation of state of a free Fermi gas --- at both zero and finite
temperatures --- see Refs.~\cite{Hua87_JW,Pat96_BH}.

To start, the Fermi wavenumber $k_{\rm F}$ is defined; $k_{\rm F}$ 
represents the momentum of the fastest moving fermion and is solely 
determined by the number density ($n\!\equiv\!N/V$) of the system, where 
$N$ is the total number of particles in our system, and $V$ is the enclosed 
volume.
 
That is,
\begin{equation}
  N=2\sum_{\bf k}\Theta(k_{\rm F}-|{\bf k}|)
   =2\int \frac{V}{(2\pi)^{3}} d^{3}k 
        \,\Theta(k_{\rm F}-|{\bf k}|)
   =V\frac{k_{\rm F}^{3}}{3\pi^{2}}\;,
 \label{FermiWN1}		  
\end{equation}
or equivalently
\begin{equation}
   k_{\rm F}=\left(3\pi^{2}n\right)^{1/3}\;.
   \label{FermiWN2}
\end{equation}
In Eq.~(\ref{FermiWN1}), $\Theta(x)$ represents the Heaviside (or step) 
function. Having defined the Fermi wavenumber $k_{\rm F}$, the energy 
density of the system is obtained from a configuration in which all 
single-particle momentum states are progressively filled in accordance 
with the Pauli exclusion principle. For a degenerate (spin-1/2) Fermi 
gas at zero temperature, exactly two fermions occupy each
single-particle state below the Fermi momentum
$p_{\rm F}\!=\!\hbar k_{\rm F}$; all remaining states above the
Fermi momentum are empty. In this manner we obtain the following
expression for the energy density:
\begin{equation}
  {\cal E} \equiv E/V = 2 \int\frac{d^{3}k}{(2\pi)^{3}}
   \Theta(k_{\rm F}-|{\bf k}|)\epsilon({\bf k}) \;,
  \label{EnergyDensity1}
\end{equation}
where $\epsilon({\bf k})$ is the single-particle energy of a 
fermion with momentum ${\bf k}$. In what follows, the most 
general free-particle dispersion (energy {\it vs}. momentum) 
relation is assumed, namely, one consistent with the postulates 
of special relativity. That is,
\begin{equation}
 \epsilon({\bf k})=\sqrt{(\hbar kc)^{2}+(mc^{2})^{2}}
                  = mc^{2}\sqrt{1+x^{2}} \;, \quad 
                  \left({\rm with} \; x\equiv 
                  \frac{\hbar kc}{mc^{2}}\right) \;.
  \label{DispRel}                           
\end{equation}
In spite of its slightly intimidating form,  the integral in
Eq.~(\ref{EnergyDensity1}) may be performed in closed form. 
We obtain 
\begin{equation}
  {\cal E} ={\cal E}_{0} \overline{{\cal E}}(\subF{x}) \;, 
  \label{EnergyDensity2} 
\end{equation}
where ${\cal E}_{0}$ is a dimensionful constant that may be written
using dimensional analysis
\begin{equation}
   {\cal E}_{0} \equiv \frac{(mc^{2})^{4}}{(\hbar c)^{3}} \;,
 \label{EnergyDensity0} 
\end{equation}
and $\overline{{\cal E}}(\subF{x}) $ is a dimensionless function of
the single variable $\subF{x}\!=\!\hbar\subF{k}c/mc^{2}$ given by
\begin{equation}
 \overline{{\cal E}}(\subF{x}) \equiv \frac{1}{\pi^{2}}
  \int_{0}^{\subF{x}} x^{2}\sqrt{1+x^{2}} \, dx=
  \frac{1}{8\pi^{2}}\left[\subF{x}\left(1+2\subF{x}^{2}\right)
  \sqrt{1+\subF{x}^{2}}-\ln\left(\subF{x}+
  \sqrt{1+\subF{x}^{2}}\,\right)\right]\;.
 \label{EBar}
\end{equation}
The pressure of the system may now be directly obtained from the
energy density by using the following thermodynamic relation --- which
is only valid at zero temperature:
\begin{equation}
 P=-\left(\frac{\partial E}{\partial V}\right)_{N,T\equiv0}
     =-\left(\frac{\partial (V{\cal E})}
       {\partial V}\right)_{N,T\equiv0}
     \equiv P_{0}\overline{P} \;. 
  \label{Pressure1}
\end{equation}  
In analogy to the energy density, dimensionful and dimensionless 
quantities for the pressure have been defined:
\begin{subequations}
\begin{eqnarray}
  &&P_{0} = {\cal E}_{0}=\frac{(mc^{2})^{4}}{(\hbar c)^{3}} \;, \\
  &&\overline{P}(\subF{x}) \equiv \left[\frac{\subF{x}}{3}
    \overline{{\cal E}}^{\prime}(\subF{x})-
    \overline{{\cal E}}(\subF{x})\right]\;.
\end{eqnarray}
\end{subequations}
It may be surprising to find that a gas of particles at
zero-temperature may still generate a non-zero pressure.  {\it It is
quantum statistics, in the form of the Pauli exclusion principle --- not
temperature --- that is responsible for generating the pressure}. It is
nevertheless surprising that quantum pressure, a purely microscopic
phenomenon, should be ultimately responsible for supporting compact
stars against gravitational collapse.

With an expression for the pressure in hand, we are finally in a
position to compute its derivative with respect to $\subF{x}$
(a quantity that we label as $\eta$). As we shall see in the next 
section, $\eta$ --- a function closely related to the zero-temperature
incompressibility --- is the only property of the EoS that Newtonian 
stars are sensitive to ~\cite{Pat96_BH}. We obtain 
\begin{equation}
 \eta \equiv \frac{dP}{dx_{\rm F}}
     = P_{0}\left[\frac{\subF{x}}{3}
        {\overline{\cal E}}^{\prime\prime}(\subF{x})
      - \frac{2}{3}\overline{{\cal E}}^{\prime}(\subF{x})\right]
     = \frac{P_{0}}{3\pi^{2}}\frac{\subF{x}^{4}}{\sqrt{1+\subF{x}^{2}}}\;.
 \label{Eta1}    
\end{equation}
The above expression has a surprisingly simple form that depends 
on the energy density only through its derivatives. Alternatively, one could 
have bypassed the above derivation in favor of the following general 
relation valid for a zero-temperature Fermi gas:
\begin{equation}
 \frac{dP}{d\subF{x}}=n\frac{d\subF{\epsilon}}{d\subF{x}}\;.    
 \label{Eta2}
\end{equation}
In view of Eq.~(\ref{Eta2}), the attentive reader may be asking why go
through the trouble of computing the energy density and the
corresponding pressure if all that is required is the dependence of
the Fermi energy on $\subF{x}$. The answer is general relativity.
While Newtonian stars depend exclusively on $\eta$, the structure of
relativistic stars (such as neutron stars) are highly sensitive to
corrections from general relativity. These corrections depend on both
the energy density and the pressure and will be treated in detail in
a future publication.

\subsection{Toy Model of White Dwarf Stars}
\label{ToyModel}
Before attempting a numerical solution to the equations of hydrostatic 
equilibrium, we consider as a warm-up exercise a toy model of a white 
dwarf star~\cite{Koon90_AW}. Assume a white dwarf star with a {\it uniform}, 
spherically symmetric mass distribution of the form
\begin{equation}
  \rho(r)=
  \begin{cases}
      \rho_{0}=3M/4\pi R^{3}\;, & {\rm if} \quad r\le R\;; \\
       0\;,                     & {\rm if} \quad r>R   \;,
  \end{cases}
 \label{UniformDensity}
\end{equation}
where $M$ and $R$ are the mass and radius of the star, respectively.
For such a spherically symmetric star, the gravitational energy 
released during the process of ``building'' the star is given by
\begin{subequations}
\begin{eqnarray}
 && E_{G} = -4\pi G \int_{0}^{R} M(r)\rho(r)r\,dr \;, \\
 && M(r) = \int_{0}^{r} 4\pi r'^{2} \rho(r')\,dr' \;.
\end{eqnarray}
\end{subequations}
For a uniform density star as assumed in Eq.~(\ref{UniformDensity}), 
it is straightforward to perform the above two integrals. Thus, 
the gravitational energy released in ``building'' such a star is 
given by
\begin{equation}
 E_{\rm G}(M,R)=-\frac{3}{5}\frac{GM^{2}}{R} \;.
 \label{Egravity}
\end{equation}
From Eq.~(\ref{Egravity}), we conclude that without a source of 
gravitational support, a star with a fixed mass $M$ will minimize its 
energy by collapsing into an object of zero radius, namely, into
a {\it black hole}.  We know, however, that white dwarf stars are 
supported  by the quantum-mechanical pressure from its degenerate 
electrons, which (at temperatures of about $10^{6}$~K) 
are fully ionized in the star (recall that $1~{\rm eV}\!\simeq\!10^{4}$~K).  
In what follows, we assume that electrons provide all the
pressure support of the star but none of its mass, while nuclei
({\it e.g.,} ${}^{4}$He, ${}^{12}$C, $\dots$) provide all the
mass but none of the pressure. The electronic contribution to
the mass of the star is inconsequential, as the ratio of electron 
to nucleon mass is approximately equal to $1:2000$.

The energy of a degenerate electron gas was computed in the
previous section. Using Eqs.~(\ref{FermiWN2}) 
and~(\ref{EnergyDensity2}) we obtain,
\begin{equation}
  E_{\rm F}(M,R)=3\pi^{2}Nm_{e}c^{2}
       \frac{{\overline{\cal E}}(\subF{x})}{\subF{x}^{3}},
 \label{EFermi}		  
\end{equation}
where $m_{e}$ is the rest mass of the electron. Naturally, the 
above expression depends on the mass and the radius of the 
star, although this dependence is implicit in $\subF{x}$.
While the toy-problem at hand is instructive of the simple,
yet subtle, physics that is displayed in compact stars, it
also serves as a useful framework to illustrate how to 
{\it scale} the equations. 


\subsubsection{Scaling the Equations}
\label{ScalingEquations}
One of the great challenges in astrophysics, and the physics of 
compact stars is certainly no exception, is the enormous range 
of scales that one must simultaneously address. For example, in 
the case of white dwarf stars it is the pressure generated by 
the degenerate electrons (constituents with a mass of 
$m_{e}\!=\!9.110\!\times\!10^{-31}$~kg) that must support stars 
with masses comparable to that of the Sun 
($M_{\odot}\!=\!1.989\!\times\!10^{30}$~kg). This represents a 
disparity in masses of 60 orders of magnitude! Without properly
{\it scaling the equations}, there is no hope of dealing with
this problem with a computer.

We start by defining $f_{\rm F}\!\equiv\!E_{\rm F}/Nm_{e}c^{2}$
from Eq.~({\ref{EFermi}), a quantity that is both dimensionless 
and intensive ({\it i.e.,} independent of the size of the system).
That is,
\begin{equation}
  \subF{f}(\subF{x})=3\pi^{2}
  \frac{{\overline{\cal E}}(\subF{x})}{\subF{x}^{3}},
 \label{fFermi}		  
\end{equation}
where a closed-form expression for ${\overline{\cal E}}(\subF{x})$ has
been displayed in Eq.~(\ref{EBar}). Note that the {\it scaled} Fermi
momentum $\subF{x}$ quantifies the importance of relativistic effects.
At low density ($\subF{x}\!\ll\!1$) the corrections from special
relativity are negligible and electrons behave as a non-relativistic
Fermi gas. In the opposite high-density limit ($\subF{x}\!\gg\!1$) the
system becomes ultra-relativistic and the ``small'' (relative to the
Fermi momentum) electron mass may be neglected. We shall see that in
the case of white dwarf stars, the most interesting physics occurs in
the $\subF{x}\!\sim\!1$ regime.

The dynamics of the star consists of a tug-of-war between gravity
that favors the collapse of the star and electron-degeneracy 
pressure that opposes the collapse. To efficiently compare these
two contributions, the contribution from gravity to the energy must
be scaled accordingly. Thus, in analogy to Eq.~(\ref{fFermi}), 
we form the corresponding dimensionless and intensive quantity for 
the gravitational energy 
($f_{\rm G}\!\equiv\!E_{\rm G}/Nm_{e}c^{2}$): 
\begin{equation}
 f_{\rm G}(M,R)=-\frac{3}{5}\left(\frac{GM}{Rc^{2}}\right)
                 \left(\frac{M}{Nm_{e}}\right)
               =-\frac{3}{5}\left(\frac{GM}{Rc^{2}}\right)
                 \left(\frac{m_{n}}{Y_{e}m_{e}}\right)\;, 
 \label{fgravity}
\end{equation}
where we have assumed that the mass of the star, $M\!=\!Am_{n}$, may 
be written exclusively in terms of its baryon number $A$ and the nucleon 
mass $m_{n}$ (the small difference between proton and neutron masses is 
neglected). This is an accurate approximation as both nuclear and 
gravitational binding energies per nucleon are small relative to the 
nucleon mass. Further, $Y_{e}\!\equiv\!Z/A$ represents the 
electron-per-baryon fraction of the star ({\it e.g.,} 
$Y_{e}\!=\!1/2$ for ${}^{4}$He and ${}^{12}$C, 
and $Y_{e}\!=\!26/56$ in the case of ${}^{56}$Fe).

The final step in the scaling procedure is to introduce dimensionful
mass $M_{0}$ and radius $R_{0}$, quantities that, when chosen wisely,
will embody the natural mass and length scales in the problem. To 
this effect we define
\begin{equation}
  \overline{M}\!\equiv\!M/M_{0} 
  \quad{\rm and}\quad
  \overline{R}\!\equiv\!R/R_{0} \;.
 \label{BarQuantities}
\end{equation}
In terms of these {\it natural} mass and length scales, the
gravitational contribution to the energy of the system takes 
the following form:
\begin{equation}
 f_{\rm G}(\overline{M},\overline{R})=-\left[\frac{3}{5}
   \left(\frac{GM_{0}}{R_{0}c^{2}}\right)
   \left(\frac{m_{n}}{Y_{e}m_{e}}\right)\right]
   \frac{\overline{M}}{\overline{R}}\;. 
 \label{fgscaled}
\end{equation}
While the dependence of the above equation on $\overline{M}$ and
$\overline{R}$ is already explicit, the Fermi gas contribution to 
the energy depends implicitly on them through $\subF{x}$. To expose 
explicitly the dependence of $f_{\rm F}$ on $\overline{M}$ and 
$\overline{R}$ we perform the following manipulation aided 
by relations derived in Sec.~\ref{EquationofState}.
\begin{equation}
  x_{\rm F}^{3}=
     \left(\frac{\hbar k_{\rm F}c}{m_{e}c^{2}}\right)^{3}=
     \left[\left(\frac{9\pi}{4}Y_{e}\right)
     \left(\frac{M_{0}}{m_{n}}\right)
     \left(\frac{\hbar c/m_{e}c^{2}}{R_{0}}
     \right)^{3}\right]
     \frac{\overline{M}}{\overline{R}^{3}}\;. 
 \label{xFscaled}
\end{equation}

We have already referred earlier to $M_{0}$ and $R_{0}$ as the 
natural mass and length scales in the problem, but their values have yet 
to be determined. Thus, they are still at our disposal. Their values
will be fixed by adopting the following choice: let the
``complicated'' expressions enclosed between square brackets in 
Eqs.~(\ref{fgscaled}) and (\ref{xFscaled}) be set equal to one.
That is,
\begin{equation}
   \left[\frac{3}{5}
   \left(\frac{GM_{0}}{R_{0}c^{2}}\right)
   \left(\frac{m_{n}}{Y_{e}m_{e}}\right)\right]=
   \left[\left(\frac{9\pi}{4}Y_{e}\right)
   \left(\frac{M_{0}}{m_{n}}\right)
   \left(\frac{\hbar c/m_{e}c^{2}}{R_{0}}
   \right)^{3}\right]=1 \;.
 \label{SettoOne}
\end{equation}
This choice implies the following values for white dwarf stars with an
electron-to-baryon ratio equal to $Y_{e}=1/2$:
\begin{subequations}
 \begin{eqnarray}
  M_{0}&=&\frac{5}{6}\sqrt{15\pi}
          \alpha_{\rm G}^{-3/2}m_{n}Y_{e}^{2}
          =10.599\,M_{\odot}\,Y_{e}^{2}
          \mathop{\longrightarrow}_{Y_{e}=1/2}
          = 2.650\,M_{\odot} \;, \\
  R_{0}&=&\frac{\sqrt{15\pi}}{2}\alpha_{\rm G}^{-1/2}
          \left(\frac{\hbar c}{m_{e}c^{2}}\right)Y_{e}
          = 17\,250~{\rm km}\,Y_{e}
          \mathop{\longrightarrow}_{Y_{e}=1/2}
          =8\,623~{\rm km}\;.
 \end{eqnarray} 
 \label{SetUnitsWD}
\end{subequations}
Here the minute dimensionless strength of the gravitational coupling 
between two nucleons has been introduced as
\begin{equation}
  \alpha_{\rm G}=\frac{Gm_{n}^{2}}{\hbar c}
                =5.906\times10^{-39} \;.
 \label{alphaG}
\end{equation}

The aim of this toy-model exercise is to find the minimum value of the
total (gravitational plus Fermi gas) energy of the star as a function
of its radius for a fixed value of its mass.  Before doing so,
however, a few comments are in order.  First, from merely scaling the
equations and with no recourse to any dynamical calculation we have
established that white dwarf stars have masses comparable to that of
our Sun but typical radii of only $10\,000$~km (recall that the radius
of the Sun is $R_{\odot}\!\approx\!700\,000$~km).  Further, we observe
that while $R_{0}$ scales with the inverse electron mass, the mass
scale $M_{0}$ is independent of it.  This suggests that neutron stars,
{\it where the neutrons provide all the pressure and all the mass},
will also have masses comparable to that of the Sun but typical radii
of only about 10~km.

Now that the necessary ``scaling'' machinery has been developed, we
return to our original toy-model problem.  Taking advantage of the 
scaling relations, the energy per electron in units of the electron 
rest energy is given by:
%
\begin{equation}
  f(\overline{M},x_{\rm F}) = 
  f_{\rm G}(\overline{M},x_{\rm F})+ f_{\rm F}(x_{\rm F})
  =-\overline{M}^{2/3}x_{\rm F}+3\pi^{2}
  \frac{{\overline{\cal E}}(\subF{x})}{\subF{x}^{3}} \;.
 \label{fTotal}
\end{equation}
The mass-radius relation of the star may now be obtained by 
demanding hydrostatic equilibrium:
\begin{equation}
 \left(\frac{\partial f(\overline{M},x_{\rm F})}
            {\partial x_{\rm F}}\right)_{\overline{M}}=0\;.
\end{equation}
While a closed-form expression has already been derived for the energy
density ${\overline{\cal E}}(\subF{x})$ in Eq.~(\ref{EBar}), it is
instructive to display explicit non-relativistic and
ultra-relativistic limits. These are given by
\begin{equation}
  f_{\rm F}(x_{\rm F})=
  \begin{cases}
    1+\frac{3}{10}x_{\rm F}^{2}  \;, & 
    {\rm if} \quad x_{\rm F}\ll 1\;; \\
    \frac{3}{4}x_{\rm F}         \;, & 
    {\rm if} \quad x_{\rm F}\gg 1\;. 
  \end{cases}
 \label{fLimits}
\end{equation}

\begin{figure}[h]
 \includegraphics{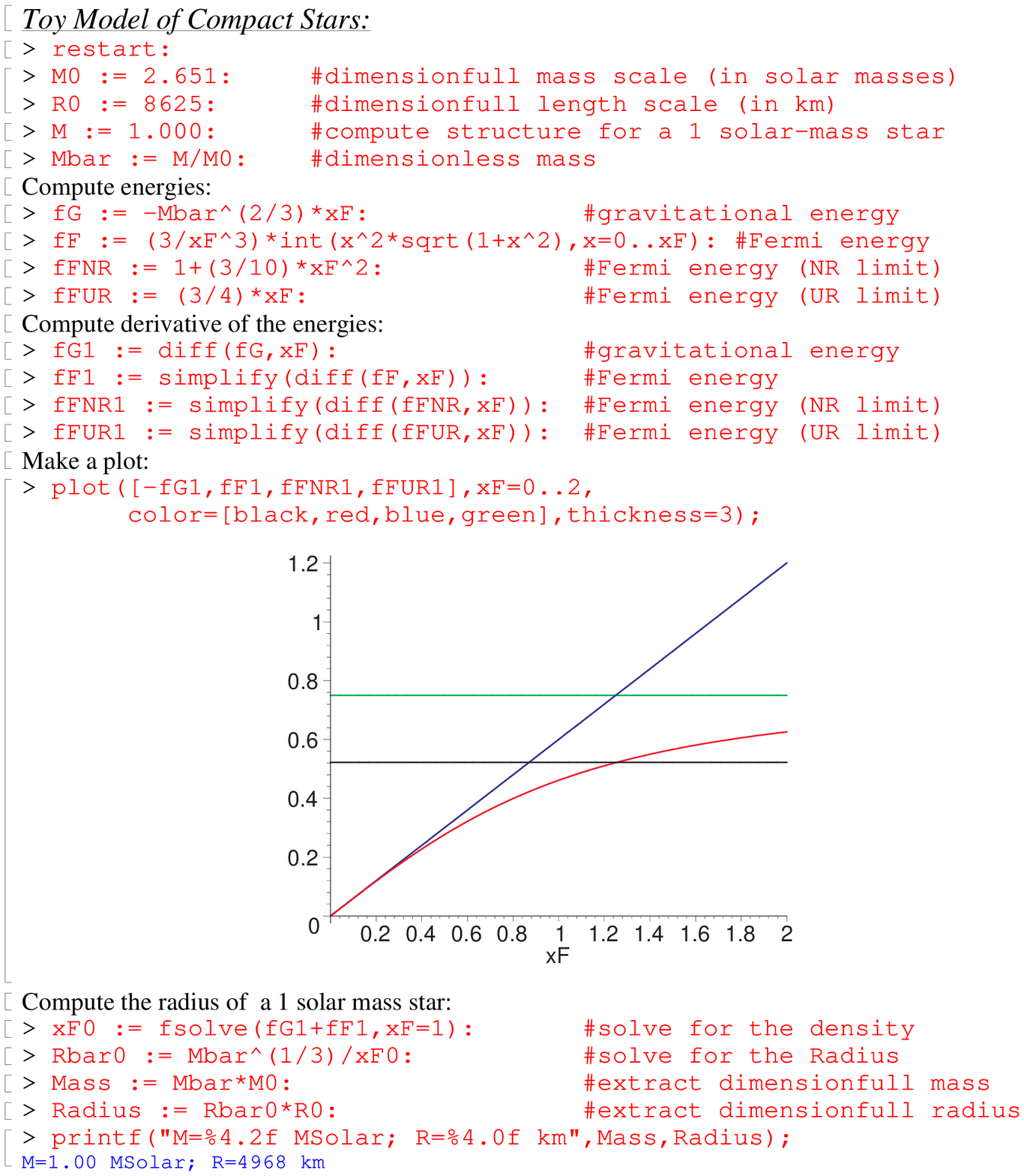}
 \caption{Maple code displaying the interplay between various 
          physical effects on the ``toy-model'' problem of a 
          $M\!=\!1M_{\odot}$ star.}
 \label{fig2}
\end{figure}

To conclude this section an output from a {\it Maple} code
has been included to illustrate how simple, within the present 
approximation, it is to compute the radius of an arbitrary mass star 
(see Fig. \ref{fig2}). For the present example, a 1 M$_{\odot}$ star has 
been used.  First, scales for input quantities, such as the dimensionful 
mass ($M_{0}$) and length ($R_{0}$), are defined.  Next, energies and their
derivatives are computed and a plot displaying the latter is
generated.  The derivative of the gravitational energy (actually the
negative of it) is constant and is displayed with a black horizontal
line.  Similarly, the derivative of the Fermi energy in the
ultra-relativistic limit (horizontal green line) is also a constant
equal to $3/4$, independent of the mass of the star.  The blue line with 
a constant slope displays the derivative of the Fermi gas energy in
the non-relativistic limit. Finally, the exact Fermi gas expression,
which interpolates between the non-relativistic and the relativistic
result, is displayed with the red line.

The equilibrium density of the star is obtained from the intersection
of the red and blue lines with the gravitational line.  In the
non-relativistic case the solution may be computed analytically to be
$x_{\rm F0}\!=\!5\overline{M}^{2/3}/3$.  However, this
non-relativistic prediction {\it overestimates} the Fermi pressure and
consequently also the radius of the star. The non-relativistic
predictions for the radius of a 1 solar-mass white dwarf star is
$R_{\rm NR}\!=\!7\,162$~km. In contrast, the result with the correct
relativistic dispersion relation is considerably smaller at an 
$R_{\rm NR}\!=\!4\,968$~km. Yet an even more dramatic discrepancy
emerges among the two models. While the non-relativistic result
guarantees the existence of an equilibrium radius for any value of the
star's mass ($\overline{R}_{\rm NR}\!=\!3/5\overline{M}^{1/3}$), the
correct dispersion relation predicts the existence of an upper limit
beyond which the pressure from the degenerate electrons can no longer
support the star against gravitational collapse. This upper mass
limit, known as the {\it Chandrasekhar mass}, is predicted in the
simple toy model to be equal to:
\begin{equation}
  M_{\rm Ch}=(3/4)^{3/2}M_{0}=1.72~M_{\odot}\;.
 \label{MChandra}
\end{equation}
As it will be shown later, accurate numerical results yield 
(for $Y_{e}\!=\!1/2$) a Chandrasekhar mass of 
$M_{\rm Ch}\!=\!1.44~M_{\odot}$. Thus, not only does the toy 
model predict the existence of a maximum mass star, but 
it does so with an 80\% accuracy.

\subsection{Numerical Analysis} 
We now return to the exact (numerical) treatment of white dwarf
stars.  While the toy model problem developed earlier provides a 
particularly simple framework to understand the interplay between 
gravity, quantum mechanics, and special relativity, a quantitative 
description of the systems demands the numerical solution of the 
hydrostatic equations [Eq.~(\ref{HydroEqs})]. For the present 
treatment, however, we continue to assume that the equation of
state is that of a simple degenerate Fermi gas. In this case the
equation of state is known analytically and it is convenient to 
incorporate it directly into the differential equation. In 
this way Eq.~(\ref{HydroEqsa}) becomes
\begin{equation}
  \frac{d\subF{x}}{dr}=-\frac{GM(r)\rho(r)}{r^2\eta}\;, 
 \label{HydroEqs2a} 
\end{equation}
where the equation of state enters only through a quantity
directly related to the zero-temperature incompressibility~\cite{Pat96_BH}. 
This quantity, $\eta\!=\!dP/d\subF{x}$, was defined and 
evaluated in Eq.~(\ref{Eta1}). Moreover, the density of the
system $\rho(r)$ is easily expressed in terms of $\subF{x}$. 
It is given by
\begin{equation}
 \rho=\left(\frac{m_{e}c^{2}}{\hbar c}\right)^{3}
      \frac{m_{n}}{3\pi^2 Y_{e}}\subF{x}^{3} \;.
 \label{Rho}
\end{equation}
At this point all necessary relations have been derived and the
equations of hydrostatic equilibrium, first displayed in
Eq.~({\ref{HydroEqs}) but with an equation of state still missing, 
may now be written in the following form:
\begin{subequations}
 \begin{eqnarray}
   &&\frac{d\subF{x}}{d\overline{r}}=-\left[
     \left(\frac{GM_{0}}{R_{0}c^{2}}\right)
     \left(\frac{m_{n}}{Y_{e}m_{e}}\right)\right]
     \frac{\overline{M}}{\overline{r}^2} 
     \frac{\sqrt{1+\subF{x}^{2}}}{\subF{x}}\;,
     \qquad \subF{x}(\overline{r}\!=\!0)\equiv
     x_{\lower2pt\hbox{$\scriptstyle{\rm Fc}$}}\;; \\
   &&\frac{d\overline{M}}{d\overline{r}}=+\left[
     \left(\frac{3\pi}{4}Y_{e}\right)
   \left(\frac{M_{0}}{m_{n}}\right)
   \left(\frac{\hbar c/m_{e}c^{2}}{R_{0}}
   \right)^{3}\right]^{-1}\overline{r}^{2}\subF{x}^{3} \;,
   \qquad\overline{M}(\overline{r}\!=\!0)\equiv 0 \;.
 \end{eqnarray}
 \label{HydroEqs2}
\end{subequations}
Here the dimensionless distance $\overline{r}$ and the central
(scaled) Fermi momentum $x_{\lower2pt\hbox{$\scriptstyle{\rm Fc}$}}$
have been introduced. The structure of the above set of differential
equations indicates that our goal of turning Eq.~({\ref{HydroEqs})
into a well-posed problem, by directly incorporating the equation of
state into the differential equations, has been accomplished. But we
have done better. By defining the natural mass and length scales of
the system ($M_{0}$ and $R_{0}$) according to Eq.~(\ref{SettoOne}),
the two long expressions in brackets in the above equations reduce to
the simple numerical values of $5/3$ and $1/3$, respectively. Finally,
then, the equations of hydrostatic equilibrium describing the
structure of white dwarf stars are given by the following expressions:
\begin{subequations}
 \begin{eqnarray}
   &&\frac{d\subF{x}}{d\overline{r}}=
    f(\overline{r};\subF{x},\overline{M}) \;, 
    \quad \subF{x}(\overline{r}\!=\!0)\equiv
    x_{\lower2pt\hbox{$\scriptstyle{\rm Fc}$}}\;; \\
   &&\frac{d\overline{M}}{d\overline{r}}=
      g(\overline{r};\subF{x},\overline{M}) \;,
    \quad\overline{M}(\overline{r}\!=\!0)\equiv 0 \;,
 \end{eqnarray} 
 \label{HydroEqsScaled}
\end{subequations}
where the two functions on the right-hand side of the equations
($f$ and $g$) are given by
\begin{equation}
     f(\overline{r};\subF{x},\overline{M})\equiv 
    -\frac{5}{3}
     \frac{\overline{M}}{\overline{r}^{2}} 
     \frac{\sqrt{1+\subF{x}^{2}}}{\subF{x}} 
     \quad{\rm and}\quad
     g(\overline{r};\subF{x},\overline{M})\equiv
    +3\,\overline{r}^{2}\subF{x}^{3} \;.
 \label{fandg}
\end{equation}
These coupled set of first-order differential equations may now be
solved using standard numerical techniques, such as the Runge-Kutta
algorithm~\cite{Koon90_AW}. However, for those students not yet
comfortable with writing their own source codes, the use of an
``off-the-shelf'' spreadsheet (here Microsoft Excel has been used),
together with a crude low-order approximation for the derivatives has
been shown to be adequate. As in the toy-model problem, solutions will
be presented using the full relativistic dispersion relation as well
as the non-relativistic approximation, where in the latter case the
square-root term appearing in the function 
$f(\overline{r};\subF{x},\overline{M})$ is set to one.

\section{Numerical Techniques for Everyone}
\label{NumericalTechniques}

In this section we present numerical solutions for the structure of
Newtonian (white-dwarf) stars by employing a variety of numerical 
techniques and programming tools. 

\subsubsection{White Dwarf Stars with Excel}
\label{Excel}

This section should be ideal for those students with a basic 
knowledge of calculus and with no programming skills. 
{\it High school} students that have learned the concept of 
derivatives in their introductory calculus class should be able 
to complete this part of the project with no problem. Indeed, one 
could use the definition of the derivative of a function $F(x)$
\begin{equation}
 F^{\prime}(x) = \lim_{h\rightarrow 0}
                 \frac{F(x+h)-F(x)}{h}\;,
 \label{Derivative1}
\end{equation}
to approximate the value of the function at a neighboring point
$x\!+\!h$. That is,
\begin{equation}
 F(x+h) = F(x)+hF^{\prime}(x)+{\cal O}(h^{2})\;. 
 \label{Derivative2}
\end{equation}
The term ${\cal O}(h^{2})$ indicates that the error that one makes
in computing the value of the function at the neighboring point
scales as the square of $h$. Thus, for this low-order approximation
of the derivative, we selected a very small value of $h$ in
order to ensure numerical accuracy. Other higher-order algorithms,
such as the venerated Runge-Kutta method, contain errors that scale
as ${\cal O}(h^{4})$. These algorithms can attain the same degree
of numerical accuracy as the one presented here with a dramatic
reduction in computational time. 

So how do we turn Eq.~(\ref{Derivative2}) to our advantage? By simply 
looking at the structure of the (scaled) equations of hydrostatic
equilibrium Eq.~(\ref{HydroEqsScaled}) we can readily write
\begin{subequations}
\begin{eqnarray}  
 \subF{x}(\overline{r}+\Delta\overline{r})   = 
          \subF{x}(\overline{r}) +
          \Delta\overline{r}\,
        f(\overline{r};\subF{x},\overline{M}) \;,
 \label{NRDis1} \\
 \overline{M}(\overline{r}+\Delta\overline{r}) = 
          \overline{M}(\overline{r}) + 
          \Delta\overline{r}\,
        g(\overline{r};\subF{x},\overline{M}) \;.
 \label{NRDis2}
\end{eqnarray}
\end{subequations}
The resulting {\it difference equations} are recursion relations that
enables one to ``leapfrog'' from point to point in a grid for which
the various points are separated by a fixed distance
$\Delta\overline{r}$.  Recursion relations such as this one are
particularly well suited to be solved with a spreadsheet.  
Fig.~\ref{WDplot} shows the results produced using Excel (lines). 

To start the solution of the difference equations one notes that the
right-hand side of the two recursion relations given above are
completely known at $\overline{r}\!=\!0$ (recall that appropriate
boundary conditions have already been specified). This enabled us to
compute both the scaled Fermi momentum and the enclosed mass on the
next grid point $\overline{r}\!=\!\Delta\overline{r}$. With this
knowledge we could again evaluate the right-hand side of the recursion
relations but now at $\overline{r}\!=\!\Delta\overline{r}$. Now the
values of $\subF{x}$ and $\overline{M}$ at the next grid point
($\overline{r}\!=\!2\Delta\overline{r}$) may be computed. We
continued in this manner until the Fermi momentum goes to zero. This
point defines the radius of the star, while the mass of the star is
the value of the enclosed mass at this last point (this point also
corresponds to when the pressure goes to zero). Up to this point, 
both radius and mass are obtained in dimensionless units. To convert 
back to physical units we simply multiplied these dimensionless 
quantities by the dimensionful parameters ($R_{0}$ and $M_{0}$) 
defined in Eq.~(\ref{SetUnitsWD}). Of course, there is no need to
repeat the full calculation for a different value of $Y_{e}$ as we 
scaled $R_{0}$ and $M_{0}$ appropriately. In order 
to create the complete mass-radius relation, we repeated the 
same procedure for a large number of central densities. 
Here a step size of $\Delta\overline{r}\!=\!0.0001$ was used
throughout and scaled central densities ranged from 0.1 to 100, moving
in steps of 0.1 at first, then to increments of 1.  Doing so produces
Fig.~\ref{WDplot} and, in particular, a Chandrasekhar limit very
close to 1.4 M$_\odot$. However, as alluded in the toy-model problem
and confirmed in this numerical calculation, there is no Chandrasekhar
limit if one uses a non-relativistic dispersion relation.
\vspace{0.4in}

\begin{figure}[H]
\begin{center}
 \includegraphics[height=8 cm]{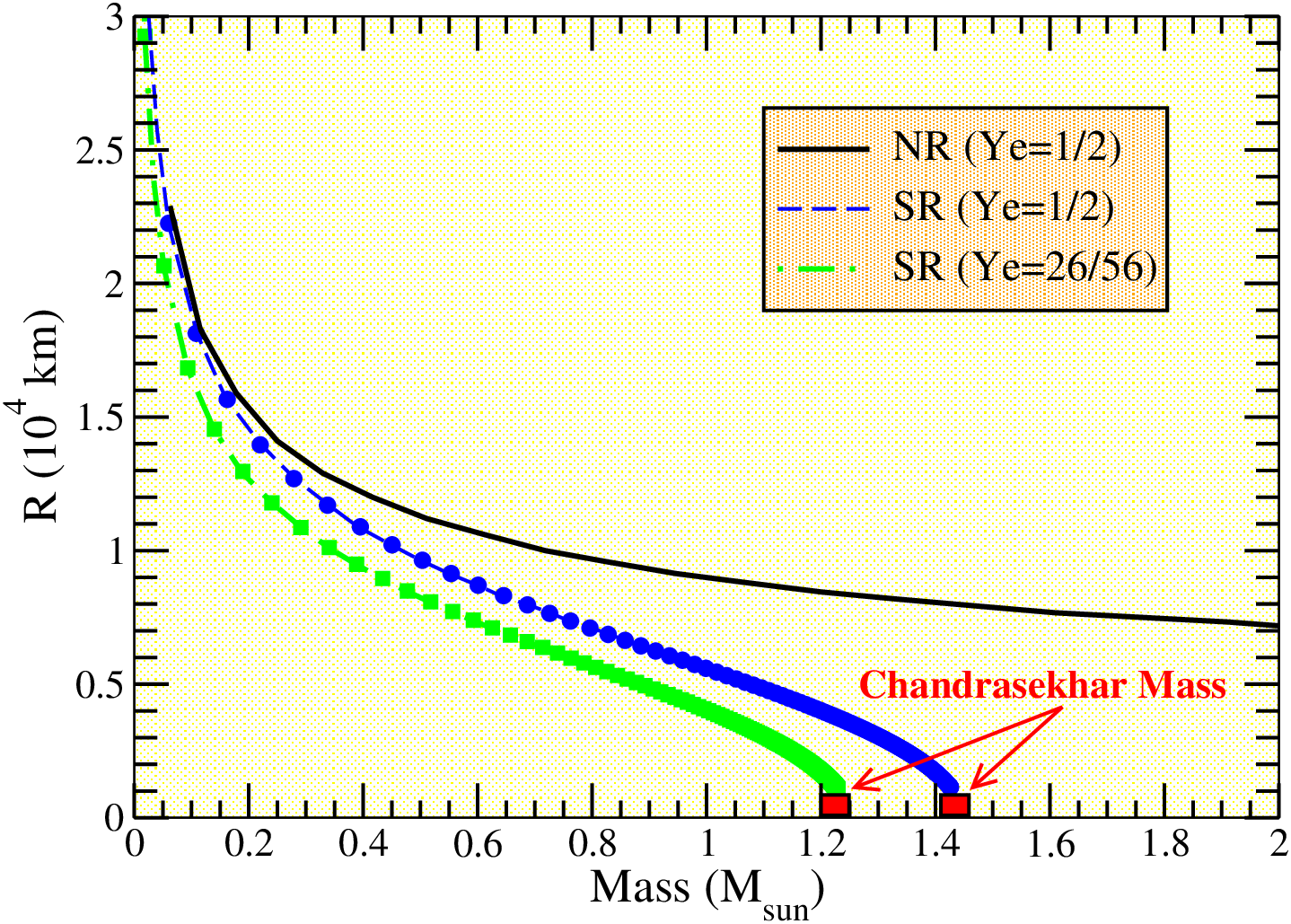}
 \caption{Mass-{\it vs}-radius relation for white dwarf stars 
          obtained using Excel (lines) and Maple (symbols).}
 \label{WDplot}
\end{center}
\end{figure}

\subsubsection{White Dwarf Stars with Maple}
\label{WDMaple}

The Maple code was created in a simple manner. It starts with a switch
(or option) that queries the user for the type of Fermi gas equation of
state (relativistic or non-relativistic) to be used.  With this
information one uses Eq.~(\ref{SetUnitsWD})
to define appropriate dimensionful mass and
radius parameters which are used after the scaled equations have been
solved.  Next, a Maple procedure was written to solve the scaled
differential equations (consistent with the switch) for a given
central (scaled) Fermi momentum using a classical numeric method.  We
noted that since in the structure equations the radius appears in the
denominator of several expressions, we could not start calculating at
a zero radius. Of course, as the limit is well defined and finite as
the radius goes to zero, one could use an extremely small value to solve 
this problem (we used a scaled radius of $10^{-23}$ for the first point).  
For a given central Fermi momentum, the equations were numerically solved 
for $\subF{x}$ and the enclosed mass $\overline{M}$ as a function of the 
scaled radius $\overline{r}$ using a very small step size of 
$\Delta\overline{r}\!=\!0.00025$.  Such a small value for 
$\Delta\overline{r}$ is necessary in order to account for the
rapidly-varying behavior of the density on the surface of the
star. The program was instructed to stop once the scaled Fermi
momentum turned negative, with the previous point defining the radius
of the star and the value of the enclosed mass at this point
representing the mass of the star. Physical dimensions were restored
by multiplying these scaled values by the dimensionful mass and radius
parameters computed earlier. Having done this once, the procedure was
repeated for a large range of central densities so one could
accurately map the mass-{\it vs}-radius relation of the star.  Once
these values were stored, a plot was generated (Fig.~\ref{WDplot}).
Fig.~\ref{WDplot} illustrates the consistency of the results using
either Excel or Maple; the data calculated using Excel are shown as
lines, and the data from Maple are shown as symbols.

\section{Summary - Concluding Remarks}
\label{Conclusions}

Isaac Newton once said: {\it If I have seen farther than others, it is
because I was standing on the shoulders of giants}.  The foundations
for the present experiment were laid by giants such as Chandrasekhar,
Fermi, Dirac, and Einstein. Their seminal work placed the fascinating
world of compact stars within the reach of the whole scientific
community. Literally, they reduced the problem of compact stellar
objects to {\it quadratures}. When these insights are combined with
the remarkable advances in computer processing and algorithms, the
problem becomes accessible to undergraduate students. (Indeed, highly
motivated high-school students should also be able to tackle most of
this problem.) The exercise reported in this manuscript is a testimony
to this fact.

Our work benefited greatly from earlier contributions by Balian and
Blaizot~\cite{Bla99_AMJ67}, and Silbar and Reddy~\cite{Sil04_AJP72}
that pioneered the idea of bringing the physics of stars to the
realm of the classroom. Here we have followed closely on their
footsteps. For the present project a group involving two upper-level
undergraduate physics students, a beginning physics graduate, and two
advanced graduate students was assembled. It was demonstrated that
with a limited knowledge of calculus and physics, undergraduate
students can easily tackle the structure equations for white dwarf
stars. Moreover, we are convinced that with a relative small amount of
mentoring, the same will be true for motivated high school
students~\cite{Sil04_AJP72}. 

Students learned several important lessons from this project. One of
them relates to the usefulness of scaling the equations--without scaling, 
the problem would have been unsolvable. This is due to
the tremendous range of scales encountered in this problem; there are
more than 60 orders of magnitude between the minute electron mass and
the immense solar mass. Another important lesson learned is that,
contrary to what seems to happen in the classroom, most problems in
physics have no analytic solution. Thus, numerical analysis is a
necessary step towards a solution. It was shown that with a
limited knowledge of calculus one can derive suitable recursion
relations to arrive at accurate solutions to the differential
equations by using a simple spreadsheet program like Microsoft Excel.
For more advanced students, symbolic programs (such as Maple or
Mathematica) provide a more efficient method for arriving at
the solutions. Due to licensing agreements, Maple was used in this
work.

In conclusion, this project successfully utilized common resources to
solve structure equations for compact objects in the Newtonian regime.
Building on this project, students are now in a position to study the
fascinating physics of neutron stars. The structure of neutron stars,
however, poses several additional challenges. First and foremost,
Newtonian gravity must be replaced by general relativity. This implies
that the structure equations must be replaced by the
Tolman-Oppenheimer-Volkoff equations~\cite{Wei72_JW,Gle00_SV,Phi02_JW}.  
Second, at the higher densities encountered in the interior of neutron
stars, the equation of state receives important corrections from the
interactions among the neutrons. That is, Pauli correlations are no
longer sufficient to describe the equation of state and some realistic
equations of state should be used~\cite{Lat01_APJ550,Sil04_AJP72}. This 
topic of intense research activity is of relevance to the physics of 
neutron stars and to the structure of those exotic compact objects 
known as hybrid and quark stars.


\begin{acknowledgments}
One of us (J.P.) is extremely grateful to the faculty and students 
of the Departament d'Estructura i Constituents de la Mat\`eria at 
the Universitat de Barcelona, especially to Profs. Centelles and 
Vi\~nas, for their hospitality during the time that this
project/lectures were developed. In addition, C.B.J. would like to thank 
Ray Kallaher for their useful discussions. Finally, we are all 
appreciative to Dr. Blessing for her help on the entirety of this work.

This work was supported in part by the U.S. Department of Energy under 
Contract No.DE-FG05-92ER40750.
\end{acknowledgments}

\vfill\eject

\bibliography{ReferencesJP}

\begin{thebibliography}{22}
\expandafter\ifx\csname natexlab\endcsname\relax\def\natexlab#1{#1}\fi
\expandafter\ifx\csname bibnamefont\endcsname\relax
  \def\bibnamefont#1{#1}\fi
\expandafter\ifx\csname bibfnamefont\endcsname\relax
  \def\bibfnamefont#1{#1}\fi
\expandafter\ifx\csname citenamefont\endcsname\relax
  \def\citenamefont#1{#1}\fi
\expandafter\ifx\csname url\endcsname\relax
  \def\url#1{\texttt{#1}}\fi
\expandafter\ifx\csname urlprefix\endcsname\relax\def\urlprefix{URL }\fi
\providecommand{\bibinfo}[2]{#2}
\providecommand{\eprint}[2][]{\url{#2}}

\bibitem[{\citenamefont{Balian and Blaizot}(1999)}]{Bla99_AMJ67}
\bibinfo{author}{\bibfnamefont{R.}~\bibnamefont{Balian}} \bibnamefont{and}
  \bibinfo{author}{\bibfnamefont{J.-P.} \bibnamefont{Blaizot}},
  \bibinfo{journal}{Am. J. Phys} \textbf{\bibinfo{volume}{67}},
  \bibinfo{pages}{1189} (\bibinfo{year}{1999}).

\bibitem[{\citenamefont{Silbar and Reddy}(2004)}]{Sil04_AJP72}
\bibinfo{author}{\bibfnamefont{R.~R.} \bibnamefont{Silbar}} \bibnamefont{and}
  \bibinfo{author}{\bibfnamefont{S.}~\bibnamefont{Reddy}},
  \bibinfo{journal}{Am. J. Phys.} \textbf{\bibinfo{volume}{72}},
  \bibinfo{pages}{892} (\bibinfo{year}{2004}),
  \eprint[http://arXiv.org/abs]{nucl-th/0309041}.

\bibitem[{\citenamefont{Fowler}(1926)}]{Fow26_RAS87}
\bibinfo{author}{\bibfnamefont{R.~H.} \bibnamefont{Fowler}},
  \bibinfo{journal}{Monthly Notices of the R. Astron. Soc.}
  \textbf{\bibinfo{volume}{87}}, \bibinfo{pages}{114} (\bibinfo{year}{1926}).

\bibitem[{\citenamefont{Chandrasekhar}(1984)}]{Cha84_RMP56}
\bibinfo{author}{\bibfnamefont{S.}~\bibnamefont{Chandrasekhar}},
  \bibinfo{journal}{Reviews of Modern Physics} \textbf{\bibinfo{volume}{56}},
  \bibinfo{pages}{137} (\bibinfo{year}{1984}).

\bibitem[{\citenamefont{Chadwick}(1932)}]{Chad32_PRSL136}
\bibinfo{author}{\bibfnamefont{J.}~\bibnamefont{Chadwick}},
  \bibinfo{journal}{Proceedings of Royal Society of London. Series A}
  \textbf{\bibinfo{volume}{136}}, \bibinfo{pages}{692} (\bibinfo{year}{1932}).

\bibitem[{\citenamefont{Baade and Zwicky}(1934)}]{Baad34_PR46}
\bibinfo{author}{\bibfnamefont{W.}~\bibnamefont{Baade}} \bibnamefont{and}
  \bibinfo{author}{\bibfnamefont{F.}~\bibnamefont{Zwicky}},
  \bibinfo{journal}{Phys. Rev.} \textbf{\bibinfo{volume}{46}},
  \bibinfo{pages}{76} (\bibinfo{year}{1934}).

\bibitem[{\citenamefont{Burnell}(2004)}]{Bel04_Sci304}
\bibinfo{author}{\bibfnamefont{S.~J.~B.} \bibnamefont{Burnell}},
  \bibinfo{journal}{Science} \textbf{\bibinfo{volume}{304}},
  \bibinfo{pages}{489} (\bibinfo{year}{2004}).

\bibitem[{\citenamefont{Hewish et~al.}(1968)\citenamefont{Hewish, Bell,
  Pilkington, Scott, and Collins}}]{Hew68_Nat217}
\bibinfo{author}{\bibfnamefont{A.}~\bibnamefont{Hewish}},
  \bibinfo{author}{\bibfnamefont{S.~J.} \bibnamefont{Bell}},
  \bibinfo{author}{\bibfnamefont{J.~D.~H.} \bibnamefont{Pilkington}},
  \bibinfo{author}{\bibfnamefont{P.~F.} \bibnamefont{Scott}}, \bibnamefont{and}
  \bibinfo{author}{\bibfnamefont{R.~A.} \bibnamefont{Collins}},
  \bibinfo{journal}{Nature} \textbf{\bibinfo{volume}{217}},
  \bibinfo{pages}{709} (\bibinfo{year}{1968}).

\bibitem[{\citenamefont{Pilkington et~al.}(1968)\citenamefont{Pilkington,
  Hewish, Bell, and Cole}}]{Pil68_Nat218}
\bibinfo{author}{\bibfnamefont{J.~D.~H.} \bibnamefont{Pilkington}},
  \bibinfo{author}{\bibfnamefont{A.}~\bibnamefont{Hewish}},
  \bibinfo{author}{\bibfnamefont{S.~J.} \bibnamefont{Bell}}, \bibnamefont{and}
  \bibinfo{author}{\bibfnamefont{T.~W.} \bibnamefont{Cole}},
  \bibinfo{journal}{Nature} \textbf{\bibinfo{volume}{218}},
  \bibinfo{pages}{126} (\bibinfo{year}{1968}).

\bibitem[{\citenamefont{Gold}(1968)}]{Gold68_Nat218}
\bibinfo{author}{\bibfnamefont{T.}~\bibnamefont{Gold}},
  \bibinfo{journal}{Nature} \textbf{\bibinfo{volume}{218}},
  \bibinfo{pages}{731} (\bibinfo{year}{1968}).

\bibitem[{\citenamefont{Gold}(1969)}]{Gold69_Nat221}
\bibinfo{author}{\bibfnamefont{T.}~\bibnamefont{Gold}},
  \bibinfo{journal}{Nature} \textbf{\bibinfo{volume}{221}}, \bibinfo{pages}{25}
  (\bibinfo{year}{1969}).

\bibitem[{Sci(2004)}]{Science304}
\bibinfo{journal}{Science} \textbf{\bibinfo{volume}{304}}
  (\bibinfo{year}{2004}).

\bibitem[{\citenamefont{Weinberg}(1972)}]{Wei72_JW}
\bibinfo{author}{\bibfnamefont{S.}~\bibnamefont{Weinberg}},
  \emph{\bibinfo{title}{Gravitation and cosmology}} (\bibinfo{publisher}{John
  Wiley \& Sons}, \bibinfo{address}{New York}, \bibinfo{year}{1972}).

\bibitem[{\citenamefont{Misner et~al.}(1973)\citenamefont{Misner, Thorne, and
  Wheeler}}]{Mis73_WH}
\bibinfo{author}{\bibfnamefont{C.~W.} \bibnamefont{Misner}},
  \bibinfo{author}{\bibfnamefont{K.}~\bibnamefont{Thorne}}, \bibnamefont{and}
  \bibinfo{author}{\bibfnamefont{J.}~\bibnamefont{Wheeler}},
  \emph{\bibinfo{title}{Gravitation}} (\bibinfo{publisher}{W. H. Freeman and
  Company}, \bibinfo{address}{New York}, \bibinfo{year}{1973}).

\bibitem[{\citenamefont{Shapiro and Teukolsky}(1983)}]{Sh83_JW}
\bibinfo{author}{\bibfnamefont{S.~L.} \bibnamefont{Shapiro}} \bibnamefont{and}
  \bibinfo{author}{\bibfnamefont{S.~A.} \bibnamefont{Teukolsky}},
  \emph{\bibinfo{title}{Black Holes, White Dwarfs, and Neutron Stars: The
  Physics of Compact Objects}} (\bibinfo{publisher}{John Wiley \& Sons},
  \bibinfo{address}{New York}, \bibinfo{year}{1983}).

\bibitem[{\citenamefont{Thorne}(1994)}]{Thor94_WWN}
\bibinfo{author}{\bibfnamefont{K.~S.} \bibnamefont{Thorne}},
  \emph{\bibinfo{title}{Black Holes and Time Warps: Einstein's Outrageous
  Legacy}} (\bibinfo{publisher}{W. W. Norton \& Company}, \bibinfo{address}{New
  York}, \bibinfo{year}{1994}), \bibinfo{edition}{2nd} ed.

\bibitem[{\citenamefont{Glendenning}(2000)}]{Gle00_SV}
\bibinfo{author}{\bibfnamefont{N.~K.} \bibnamefont{Glendenning}},
  \emph{\bibinfo{title}{Compact Stars}} (\bibinfo{publisher}{Springer-Verlag},
  \bibinfo{address}{New York, Berlin, Heidelberg}, \bibinfo{year}{2000}),
  \bibinfo{edition}{2nd} ed.

\bibitem[{\citenamefont{Phillips}(2002)}]{Phi02_JW}
\bibinfo{author}{\bibfnamefont{A.~C.} \bibnamefont{Phillips}},
  \emph{\bibinfo{title}{The Physics of Stars}} (\bibinfo{publisher}{John Wiley
  \& Sons}, \bibinfo{address}{New York}, \bibinfo{year}{2002}),
  \bibinfo{edition}{2nd} ed.

\bibitem[{\citenamefont{Huang}(1987)}]{Hua87_JW}
\bibinfo{author}{\bibfnamefont{K.}~\bibnamefont{Huang}},
  \emph{\bibinfo{title}{Statistical Mechanics}} (\bibinfo{publisher}{John Wiley
  \& Sons}, \bibinfo{year}{1987}), \bibinfo{edition}{2nd} ed.

\bibitem[{\citenamefont{Pathria}(1996)}]{Pat96_BH}
\bibinfo{author}{\bibfnamefont{R.~K.} \bibnamefont{Pathria}},
  \emph{\bibinfo{title}{Statistical Mechanics}}
  (\bibinfo{publisher}{Butterworth-Heinemann}, \bibinfo{year}{1996}),
  \bibinfo{edition}{2nd} ed.

\bibitem[{\citenamefont{Koonin and Meredith}(1990)}]{Koon90_AW}
\bibinfo{author}{\bibfnamefont{S.~E.} \bibnamefont{Koonin}} \bibnamefont{and}
  \bibinfo{author}{\bibfnamefont{D.~C.} \bibnamefont{Meredith}},
  \emph{\bibinfo{title}{Computational Physics}}
  (\bibinfo{publisher}{Addison-Wesley}, \bibinfo{address}{Reading, MA},
  \bibinfo{year}{1990}).

\bibitem[{\citenamefont{Lattimer and Prakash}(2001)}]{Lat01_APJ550}
\bibinfo{author}{\bibfnamefont{J.~M.} \bibnamefont{Lattimer}} \bibnamefont{and}
  \bibinfo{author}{\bibfnamefont{M.}~\bibnamefont{Prakash}},
  \bibinfo{journal}{ApJ} \textbf{\bibinfo{volume}{550}}, \bibinfo{pages}{426}
  (\bibinfo{year}{2001}), \eprint[http://arXiv.org/abs]{astro-ph/0002232}.

\end{thebibliography}

\end{document}